\documentstyle[12pt,epsf]{article}

\def\thefootnote{\fnsymbol{footnote}}

\newcommand{\eq}{\begin{equation}}
\newcommand{\en}{\end{equation}}
\newcommand{\eqa}{\begin{eqnarray}}
\newcommand{\ena}{\end{eqnarray}}
\newcommand{\bea}{\begin{eqnarray}}
\newcommand{\eea}{\end{eqnarray}}

\newcommand{\br}{\langle}
\newcommand{\kt}{\rangle}

\begin{document}
\begin{titlepage}
\vskip0.5cm
\begin{flushright}
DFTT 10/02\\
DESY 02-042\\
\end{flushright}
\vskip0.5cm
\begin{center}
{\Large\bf Finite temperature corrections in 2d integrable models}
\end{center}
\vskip 1.3cm
\centerline{
M. Caselle$^a$\footnote{e--mail: caselle~@to.infn.it}
 and M. Hasenbusch$^b$\footnote{e--mail: Martin.Hasenbusch@desy.de}}
 \vskip 1.0cm
 \centerline{\sl  $^a$ Dipartimento di Fisica
 Teorica dell'Universit\`a di Torino}
 \centerline{\sl Istituto Nazionale di Fisica Nucleare, Sezione di Torino}
 \centerline{\sl via P.Giuria 1, I-10125 Torino, Italy}
 \vskip .4 cm
 \centerline{\sl $^b$ NIC/DESY Zeuthen, Platanenallee 6, 
 D-15738 Zeuthen, Germany}
 \vskip 1.cm

\begin{abstract}
We study the finite size corrections for the magnetization and the internal
energy of the 2d Ising model in a magnetic field by using transfer matrix
techniques. We compare these corrections
with the functional form recently proposed by Delfino and LeClair-Mussardo for
the finite temperature behaviour of one-point functions in integrable 2d quantum
field theories.
We find a perfect agreement between theoretical expectations and numerical
results. Assuming the proposed functional form as an input in our analysis
we obtain a relevant improvement in the precision of the
 continuum limit estimates of both quantities.

\end{abstract}
\end{titlepage}

\setcounter{footnote}{0}
\def\thefootnote{\arabic{footnote}}

\section{Introduction}
Despite the impressive progress of these last years, 
2d quantum field theories (QFT)
 still provide many interesting open problems. One of
these is the finite temperature behaviour of correlators. Besides its
experimental and theoretical relevance, this issue is particularly 
important when these theories are studied by means of numerical simulations. In
fact
  choosing a finite temperature
 setting corresponds,
in the euclidean formulation of the theory,
 to compactify the (imaginary) time direction on a circle whose circumference
 $R$ coincides with the inverse  temperature. Thus 
the finite temperature corrections become, in this framework,
 finite size corrections in the cylinder geometry which, as it is well known,
 play a crucial role in the extraction of continuum limit expectation values
 from numerical simulations.

 While for a generic 2d QFT finding a functional form for the finite temperature
correction seems a hopeless task,
 recently two interesting proposals by Delfino~\cite{d2001} and LeClair and
 Mussardo~\cite{lm99} (based on a previous work by LeClair and
 collaborators~\cite{llss96})
 appeared in the literature
 to address this problem
in the simpler case of integrable models.
In particular, in~\cite{d2001,lm99}, the authors studied the finite temperature
behaviour of one-point correlation functions.
General arguments suggest that, outside the critical point, any one-point
function evaluated on a finite size lattice $\langle\Phi\rangle_R$
  should approach its infinite volume
limit$\langle\Phi_l\rangle_{R=\infty}$ 
with an exponential decay of the type
\eq
\langle\Phi_l\rangle_R\sim\langle\Phi_l\rangle_{R=\infty} + C e^{-\frac{R}{\xi}}
+\cdots
\label{eq1}
\en
  $\xi$ being the 
 correlation length (i.e. the inverse of the 
 lowest mass of the theory). However with these
 considerations nothing can be said
 on the constant $C$ and the higher order terms in the above
 equation.

The main achievement of~\cite{d2001,lm99} was to show that the constant $C$ is
indeed universal and to predict its value. At the same time they were able to
give an explicit expression for higher order corrections.

The two proposals~\cite{d2001} and~\cite{lm99}, 
have different theoretical starting points and indeed the
predictions for these higher order corrections turn out to be different in the
two approaches
 (for a critical comparison and a  discussion of
these differences see~\cite{m2001}). However, if one looks at the first few
orders in a low temperature expansion (those which do not involve multi-particle
form factors) the results coincide.

The aim of this paper is to test these proposals in a particular integrable 
model: the 2d Ising model in a magnetic field. This model has recently
 been the subject of several theoretical and numerical
 studies~\cite{fz2001}-\cite{cgm2001}.
 In particular,
 in the numerical analyses (performed using
 transfer matrix techniques~\cite{ch99} or
 montecarlo simulations~\cite{cgm99,cgm2001}) 
a great attention was devoted to the
 analysis of finite size correction. However, despite this careful
 treatment these correction were the major source of uncertainty in the final
 results. Our goal in this paper is thus twofold: 
first we shall use the numerical
 results for finite size lattices to test the proposals of ~\cite{d2001,lm99}.
 Second, we shall see if, assuming the functional form for the finite size
 corrections proposed in~\cite{d2001,lm99} we can improve the precision of the
 continuum limit results for the one -point correlation functions.

This paper is organized as follows. In sect.2 we shall briefly discuss the
model, the observables and the proposal of~\cite{d2001,lm99}.
 Sect.3 will be
devoted to a discussion of our numerical test and to a
comparison with~\cite{d2001,lm99}. Finally sect.4 will be devoted to 
some concluding remarks.

\section{Ising model in a magnetic field}
The  Ising model in a magnetic field is defined
by the partition function
\eq
Z=\sum_{\sigma_i=\pm1}e^{\beta(\sum_{\br n,m \kt}\sigma_n\sigma_m
+H\sum_n\sigma_n)}
\label{zz1}
\en
where the field variable $\sigma_n$ takes the values $\{\pm 1\}$;
$n\equiv(n_0,n_1)$ labels the sites of a square lattice  size $L_0$ and $L_1$
in the two directions and lattice spacing $a$\footnote{Since the lattice spacing
will play no role in the following we shall set $a=1$ in the rest of the 
paper.}.
 $\br n,m \kt$ 
denotes nearest neighbour sites on the lattice.
In the following 
we shall treat asymmetrically the two directions. We shall denote
$n_0$ as the compactified  ``time'' coordinate and $n_1$ as the space one.
The number of sites  of
the lattice will be denoted by  $N\equiv L_0 L_1$. The lattice extent in the
transverse direction will be denoted as $R\equiv L_0$. This length will play a
major role in the following. In fact our goal will be to
describe the $R$ dependence of the expectation values of the spin and energy 
operators of the theory.

In order to select only the magnetic perturbation, the coupling
 $\beta$ must be fixed to its
critical value
 $$\beta=\beta_c=\frac12\log{(\sqrt{2}+1)}=0.4406868...$$
 by defining  $h_l=\beta_c H$ we end up with 
\eq
Z(h_l)=\sum_{\sigma_i=\pm1}e^{\beta_c\sum_{\br n,m \kt }\sigma_m\sigma_m
+h_l\sum_n\sigma_n}   \;\;\; .
\label{fupart}
\en

In the continuum limit the model is described
by the action:
\eq
{\cal A} = {\cal A}_0 + h \int d^2x \, \sigma(x) \,\,\,.
\label{action2}
\en
where $\sigma(x)$ is the perturbing operator and the perturbing field is the
magnetic field $h$.

In this paper we want to use results obtained in the {\sl continuum} theory to
describe the finite size scaling behaviour of the {\sl lattice} model. In
general this would require a precise definition of the relations between lattice
and continuum quantities like for instance that between $h_l$ which appears in
eq.(\ref{fupart}) and the perturbing field $h$. However, as we shall see below,
we shall only be interested in adimensional ratios in which all these
normalizations cancel out. Thus we shall neglect this problem in the following.
The interested reader can find a careful treatment of this issue in~\cite{ch99}.

\subsection{Lattice operators}
As a first step of our analysis 
let us define the lattice analogous of the spin and energy
operators of the continuum theory\footnote{Strictly speaking these 
lattice operators do not correspond
 exactly to the continuum ones but are instead
 linear combinations
of all possible relevant and irrelevant operators of the continuum theory 
with compatible
symmetry properties with respect to the $Z_2$ symmetry of the model
(odd for the spin operator and even for the energy one). However near 
the critical point
this linear combination is dominated by the corresponding
 relevant operator ($\sigma$ for $\sigma_l$ and $\epsilon$ for $
\epsilon_l$) and the only
remaining freedom will be a conversion constant relating the continuum and 
lattice versions of the two operators. As mentioned above we shall neglect in
the following these normalization constants.}.
  The simplest choices for these lattice analogous are
\begin{itemize}
\item Spin operator
\eq
\sigma_l(x)\equiv \sigma_x
\en
i.e. the operator which associates to each site of the lattice the value of the
spin in that site. 
\item Energy operator
\eq
\epsilon_l(x)\equiv \frac14\sigma_x\left(\sum_{y~n.n.~x} \sigma_y\right)
-\epsilon_b  % -\epsilon_{h=0}
\label{defene}
\en
where the sum runs over  the four nearest neighbour sites $y$  of $x$.
$\epsilon_b$ represents a constant ``bulk'' term which we shall discuss below
\footnote{As a matter of fact, for technical reasons, in our Transfer Matrix 
calculations
we evaluated only the ``time-like'' part of the action i.e. $\sum_{n_0,
n_1}\sigma_{(n_0,n_1)}\sigma_{(n_0+1,n_1)}$. While, obviously, this choice
 makes no difference in the thermodynamic limit, it might have some effect for
 finite values of $R$. We shall further discuss this point in sect.3.1 below.}.

\end{itemize}
The index $l$ indicates that these are the lattice discretizations of the
continuous operators.
We shall denote in the following the normalized sum over all the sites of these
operators simply as
\eq
\sigma_l\equiv\frac1N\sum_x\sigma_l(x) \hskip2cm
\epsilon_l\equiv\frac1N\sum_x\epsilon_l(x)   \;\;\; .
\en

In the following we shall be interested in the expectation value of these
lattice operators. More precisely we shall be interested in:

\begin{itemize}

\item {\bf Magnetization}

 The magnetization per site $M(h_l)$ defined as
\eq
M(h_l)\equiv \frac1N \frac{\partial}
{\partial h_l}(\log~Z(h_l))\vert_{\beta=\beta_c}
=  \frac1N \br\sum_i \sigma_i \kt \;\;\; .
\en
which implies
\eq
M(h_l)=\br\sigma_l\kt  \;\;\; .
\en

\item {\bf Internal Energy}

The internal energy density $\hat E(h_l)$ defined as
\eq
\hat E(h_l) 
\equiv \frac{1}{2N}\br \sum_{\br n,m \kt}\sigma_n\sigma_m \kt   \;\;\; .
\en
It is important to stress that
 in this case one also has to take into account
 a bulk analytic contribution (as it happens also for the free energy itself)
$E_b(h_l)$ which is an even function of $h_l$.
Let us define $\epsilon_b\equiv E_b(0)$.
The value of $E_b(0)$ can be easily evaluated (for instance by
using Kramers-Wannier duality) to be $\epsilon_b=\frac{1}{\sqrt2}$. 
Let us define $E(h_l)\equiv \hat E(h_l)-\epsilon_b$, we have
\eq
E(h_l) = \frac{1}{2N} \br \sum_{\br n,m\kt}\sigma_n\sigma_m
\kt
-\frac{1}{\sqrt{2}}   \;\;\; .
\label{bulk}
\en
Hence we have 
\eq
E(h_l)=\br\epsilon_l\kt   \;\;\; .
\en

\end{itemize}

In the following we shall in particular be interested in the ratios
\eq
\frac{\langle\sigma_l\rangle_R}{\,\,\,\,\langle\sigma_l\rangle_{R=\infty}}
\hskip2cm
\frac{\langle\epsilon_l\rangle_R}{\,\,\,\,\langle\epsilon_l\rangle_{R=\infty}}
\en

$\langle\Phi_l\rangle_R$ being the mean value on a lattice with transverse
extent $R$ of the lattice operator $\Phi_l$. Since in the ratio all the
normalization constants cancel out, we can identify the two ratios with the
analogous ones {\sl evaluated in the continuum theory, with the continuum
operators}.
\eq
\frac{\langle\Phi_l\rangle_R}{\,\,\,\,\langle\Phi_l\rangle_{R=\infty}}~~=~~
\frac{\langle\Phi\rangle_R}{\,\,\,\,\langle\Phi\rangle_{R=\infty}}~~.
\en
Notice that to perform this identification it is mandatory to eliminate the bulk
correction as we did in eq.(\ref{bulk})

\subsection{Critical behaviour}
The critical behaviour of magnetization and internal energy
 can be easily obtained by
means of standard renormalization group methods. 
One finds
\eq
M\propto |h|^{d/y_h-1}
\label{asy2}
\en
\eq
E\propto |h|^{(d-y_t)/y_h}  \;\;\; .
\label{asy4}
\en
where $y_h$, $y_t$ are the are the magnetic and thermal RG-exponents
respectively.

From the exact solution of the Ising model at the critical point we know that
$y_h=\frac{15}{8}$ 
and $y_t=1$. Inserting these values in the above expressions we
find

\eq
M\propto |h|^{\frac{1}{15}}
\label{asy2b}
\en
\eq
E\propto |h|^{\frac{8}{15}}  \;\;\; .
\label{asy4b}
\en

From the conformal field theory (CFT) description of the Ising model at the
critical point it is possible to construct also the $h$ dependence of the
following terms in the two scaling functions (\ref{asy2b}) and (\ref{asy4b}).
A detailed analysis can be found in~\cite{ch99}. We shall make use of this
result in the following.

\subsection{S-matrix results}

 In 1989 A. Zamolodchikov~\cite{z89} suggested  that
 the scaling limit of the Ising Model in a
magnetic field  could be described by a
a scattering theory which 
contains eight different species of self-conjugated particles
$A_{a}$, $a=1,\ldots,8$ with masses
\eqa
m_2 &=& 2 m_1 \cos\frac{\pi}{5} = (1.6180339887..) \,m_1\,,\nonumber\\
m_3 &=& 2 m_1 \cos\frac{\pi}{30} = (1.9890437907..) \,m_1\,,\nonumber\\
m_4 &=& 2 m_2 \cos\frac{7\pi}{30} = (2.4048671724..) \,m_1\,,\nonumber \\
m_5 &=& 2 m_2 \cos\frac{2\pi}{15} = (2.9562952015..) \,m_1\,,\\
m_6 &=& 2 m_2 \cos\frac{\pi}{30} = (3.2183404585..) \,m_1\,,\nonumber\\
m_7 &=& 4 m_2 \cos\frac{\pi}{5}\cos\frac{7\pi}{30} = (3.8911568233..) \,m_1\,,
\nonumber\\
m_8 &=& 4 m_2 \cos\frac{\pi}{5}\cos\frac{2\pi}{15} = (4.7833861168..) \,m_1\,,
\nonumber
\ena
where $m_1(h)$ is the lowest mass of the theory. 
which coincides with
the inverse of the (exponential) correlation length. After Zamolodchikov's 
paper several other interesting
results were obtained, ranging from the explicit values of various critical
amplitudes~\cite{f93,flzz97,d97}
 to the values of the overlap amplitudes of correlators~\cite{dm95,ds96}. 
All these
predictions have been tested with numerical simulations both in the 1d Ising
quantum chain~\cite{hs89}, in the dilute $A_3$ IRF 
(Interaction Round a Face) model~\cite{irf}
 (which is another realization of the scaling
Ising model in a magnetic field) and directly in the 2d lattice Ising 
model~\cite{ch99,cgm99,cgm2001,chpv00,chpv01}
 and in all cases a full
agreement between S-matrix predictions and numerical results was found.

\subsection{Delfino and LeClair-Mussardo proposals}
One of the most interesting features of the approach discussed 
in~\cite{d2001} is that, by using the form-factor technology the author was able
to give a very explicit and compact expression for the first few orders of the
finite size corrections both for $\langle\sigma\rangle$ and
 $\langle\epsilon\rangle$ for the Ising model in a magnetic field. 
   This result will be of great importance for our
 analysis.
 
According to~\cite{d2001} 
a generic one point function $\langle\Phi\rangle_R$ evaluated on a
cylinder of transverse size $R$ approaches exponentially its asymptotic value
$\langle\Phi\rangle_{R=\infty}$ with the following law.
\eq
\frac{\langle\Phi\rangle_R}{\,\,\,\,\langle\Phi\rangle_{R=\infty}}=1+
\frac{1}{\pi}\sum_{i=1}^3 A_i^\Phi\,K_0(m_iR)+O(e^{-2m_1R})\,,
\label{universal}
\en
where $m_1,m_2$ and $m_3$
are the first three masses (the only ones below the lowest
 pair creation threshold)
of the Zamolodchikov's solution discussed above
 and $K_0$ denotes the zeroth order modified Bessel
function.

The major result of~\cite{d2001}  was to show that 
the $A_i^\Phi$ constants are indeed universal and can be evaluated exactly in
the framework of the S-matrix description of the model. Their values in the case
of the two operators in which we are interested here are given in tab.
\ref{tab1}. As mentioned in the introduction these results turn out to agree at
this order with those obtained by LeClair and Mussardo in~\cite{lm99}.

\begin{table}[h]
\caption{\sl
The universal amplitudes entering the expansion
(\ref{universal}) for the Ising model in a magnetic field.}
\vskip 0.2cm
%\begin{center}
\begin{tabular}{|c||c|c|}\hline
$\Phi$      & $\sigma$ & $\varepsilon$ \\ \hline
$ A^\Phi_1$ & $-8.0999744..$ & $-17.893304..$      \\
$ A^\Phi_2$ & $-21.206008..$ & $-24.946727..$      \\
$ A^\Phi_3$ & $-32.045891..$ & $-53.679951..$      \\ \hline
%\end{center}
\end{tabular}
\label{tab1}
\end{table}

\section{Numerical test}
In order to test the above prediction we performed a numerical study of the
finite size corrections by using the results of the transfer matrix analysis
discussed in~\cite{ch99} to which we refer for further
 details on the algorithm and on the raw data. We only recall here some general
 informations on the experiment.
We studied the Ising model for the 19 different values of the external magnetic
field listed in tab.\ref{tab2}. For each choice of $h$ we studied lattices with
values of $R$ ranging from 10 to 21. For each values of $h$ and each lattice
size we evaluated the mean value of the magnetization and internal energy (see
however the footnote below eq.(\ref{defene})). We
show in tab.\ref{tab2bis} a typical sample of our data. An important ingredient
of eq.(\ref{universal}) are the values of the first three masses of the theory.
In principle these masses could be evaluated directly from the S-matrix
solution, using the suitable normalization constants to match with the lattice
discretization and then imposing the standard $h$ dependence dictated by the RG
analysis. However
it is important to stress that at the level of precision of our analysis,
these three masses show rather large corrections to
scaling for the values of $h$ that we study. 
To avoid large systematic errors in the analysis
of the finite size correction it is mandatory to take into account these
corrections (they are discussed in great detail in~\cite{ch99}). An 
 alternative route, which turns out to be much simpler,
 is to evaluate these
 masses, when possible, directly from the  transfer matrix 
calculation. For the present analysis we used this second route.
 We report for completeness these values in tab.\ref{tab2}
 and refer to~\cite{ch99} for a detailed discussion of this
table.

\begin{table}[h]
\caption{\sl The first three masses.}
\vskip 0.2cm
\begin{tabular}{|l|l|l|l|l|}
\hline
   $  h$            &   $1/m_1$         &   $1/m_2$   &  $1/m_3$        \\
\hline
0.20              & 0.59778522553(1)   & 0.37795775263(1)    & 0.310888(1) \\
0.19              & 0.61388448719(1)   & 0.38765653507(1)    & 0.318578(1) \\
0.18              & 0.63134670477(1)   & 0.39818995529(1)    & 0.326940(1) \\
0.17              & 0.65037325706(1)   & 0.40968266918(1) & 0.336077(2) \\
0.16              & 0.67120940172(1)   & 0.42228634593(5) & 0.346115(3) \\
0.15              & 0.69415734924(1)   & 0.43618773124(1) & 0.357209(3) \\
0.14              & 0.71959442645(1)   & 0.45161985381(4) & 0.369548(4) \\
0.13              & 0.74799884641(1)   & 0.4688779288(2)  & 0.38338(1) \\
0.12              & 0.77998715416(1)& 0.488342470(1)   & 0.3990(1) \\
0.11              & 0.81637015277(1)& 0.510513817(1)   & 0.4168(1) \\
0.10              & 0.85823913569(5)& 0.5360654(1)     & 0.4374(5) \\
0.09              & 0.9071039295(1) & 0.5659287(6)     & 0.4624(5) \\
0.08              & 0.965123997(1)  & 0.60144(1)       & 0.492(1) \\
0.075             & 0.998514180(1)  & 0.62189(1)       & 0.508(1) \\
0.066103019026467 & 1.067300500(2)  & 0.66405(5)       & 0.543(1) \\
0.055085849188723 & 1.17524158(3)   & 0.7305(1)        &      \\
0.044068679350978 & 1.322589(6)     & 0.82(1)          &   \\
0.033051509513233 & 1.54057(2)      &                 &  \\
0.022034339675489 & 1.91(1)         &                &  \\
\hline
\end{tabular}
\label{tab2}
\end{table}

\begin{table}[h]
\caption{\sl Magnetization and Internal Energy at $h=0.075$}
\vskip 0.2cm
\begin{tabular}{|l|l|l|}
\hline
   $  R$            &   $M$         &   $E$         \\
\hline
 10 &   0.884069072534 & 0.122792065846   \\
 11  &  0.884096308760 & 0.122797848782   \\
 12   & 0.884105802401 & 0.122799890318   \\
 13  &  0.884109130761 & 0.122800613003   \\
 14  &  0.884110302840 & 0.122800869443   \\
 15  &  0.884110717053 & 0.122800960633   \\
 16  &  0.884110863864 & 0.122800993122   \\
 17  &  0.884110916027 & 0.122801004716   \\
 18  &  0.884110934601 & 0.122801008861   \\
 19  &  0.884110941226 & 0.122801010344   \\
 20  &  0.884110943594 & 0.122801010875   \\
 21  &  0.884110944441 & 0.122801023168   \\
\hline
\end{tabular}
\label{tab2bis}
\end{table}

\subsection{Analysis of the data}

 We performed a two steps analysis
\begin{description}
\item{1]}
First of all, for each value of $h$ we fitted the values of
 $<\sigma>$ and 
 $<\epsilon>$ as a function of $R$, 
 according to eq.(\ref{universal}):
The results are reported in tab.\ref{tab3}.
Here are some comments on the fits:
\begin{itemize}
\item
We always kept the asymptotic value of 
$<\sigma>$ and 
 $<\epsilon>$ as a free parameter.
\item
The error of the input data was always of the order of $10^{-12}$. 
We started by fitting all the values of $R$ and then eliminated them one by one,
starting form the smallest ones, until we reached a reduced $\chi^2$
 equal or smaller than 1. Then  
we accepted the result of the fit and stopped the analysis. 
 These acceptable $\chi^2$'s
 could be achieved for $h\geq0.075$  keeping only the first
 term in eq.(\ref{universal}). For $h=0.066..,0.055..,0.044..$ we had to
 introduce also the second mass.
 and finally for $h=0.033..,0.022..$ all the three terms were 
 needed\footnote{This trend
 is due to the fact that we studied, for each value of $h$, the same range of
values of $R$ while  $m_1$ decreases as $h$ goes to $0$. This means that 
 the
argument of the first Bessel function  in eq.(\ref{universal}) (i.e. $m_1R$) 
 decreases 
as $h\to0$ thus allowing to detect, within the errors of our data 
 also higher order terms of the equation.}. 
 The values of the masses which could not be directly evaluated via transfer
 matrix (the empty spaces in tab.\ref{tab2}) were evaluated using the S-matrix
 results and keeping into account the proper corrections to scaling.
\item
For all the values of $h$ the asymptotic values that we
obtained for $<\sigma>$ and 
 $<\epsilon>$ from these fits are compatible within the errors with those 
  quoted in~\cite{ch99}. In general the present estimates are more
 precise. This gain in precision becomes larger and larger as $h$ decreases.
For instance for $h=0.075$ we found 
\eq
E=0.122801011162(5)\hskip2cm
M=0.884110944919(2)
\label{comp1}
\en
to be compared with the values reported in~\cite{ch99}
\eq
E=0.1228010112(1)\hskip2cm
M=0.88411094491(1)
\label{comp2}
\en
with an improvement of more than one order of magnitude 
in precision.

Let us stress that both this improvement and the fact that {\bf all} our results
agree within the errors (which means typically an agreement within
 10 significative digits as in eq.s(\ref{comp1}) and (\ref{comp2}) ) 
with those of ref~\cite{ch99}  are rather impressive
evidences of the correctness of eq.(\ref{universal})\footnote{The results
of~\cite{ch99} were obtained with an  iterative
method (see sect.5.1 of ref.\cite{ch99} and sect.3.1.2 of \cite{chpv00}
 for a discussion) based on the general
assumptions on the finite size behaviour of  one point functions summarized in
eq.(\ref{eq1}). In particular the discussion of sect.3.1.2 of \cite{chpv00}
shows that the agreement
with the present results is not a case but is due to the fact 
that the iterative method actually
 mimics the exact behaviour of eq.(\ref{universal}). 
It would be nice to see if this iterative
algorithm keeps its predicting power even  when the theory is not
integrable and no exact prediction can be deduced from a S-matrix
analysis.}. In fact, as it can be seen
by looking at tab.\ref{tab2bis}, the finite size corrections which eq.
(\ref{universal}) is able to describe are much larger 
(five order of magnitude!) than the uncertainties of
the data: the corrections are of the order of $10^{-7}$ while the precision of
the input data is $10^{-12}$.
\item
For all the values of $h$ we obtained stable and reliable values for
 $A_1^{\Phi}$. These values are reported in tab.\ref{tab3}. The same was not 
true for  $A_2^{\Phi}$. This is not strange. In all the similar analyses
performed in~\cite{ch99} it was impossible to have reliable estimates for
the amplitudes of the subleading exponents due to the large systematic
deviations induced by the uncertainties in the leading corrections.

\end{itemize}

Looking at the data of tab.\ref{tab3} one can see that there are rather
large corrections to scaling. This is true in particular for 
$A_1^{\epsilon}$ (for which it 
is possible that the rather large magnitude of the correction 
 is a consequence of our
asymmetric definition for the internal energy, see footnote below eq.
(\ref{defene})). In any case it is clear that in 
order to test the predicted values for 
$A_1^{\sigma}$ and $A_1^{\epsilon}$ it is mandatory to discuss these deviations.
To this end we performed a second level of analysis

\item{2]}
The aim of this second level of analysis is 
to use the data reported in tab.\ref{tab3} to reach the scaling limit
values for  $A_1^{\sigma}$ and $A_1^{\epsilon}$. This problem is exactly the
same that we had to face in~\cite{ch99} to extract the continuum limit values of
the  critical amplitudes and thus we use here  the same method developed and
discussed in  sect.6 of~\cite{ch99} to which we refer the interested reader for
further details.

We used as fitting functions
\eq
A_1^{\sigma}(h_l)=A_1^{\sigma}(1+
b_{1,\sigma}|h_l|^{\frac{16}{15}}+
b_{2,\sigma}|h_l|^{\frac{22}{15}})
\en
\eq
A_1^{\epsilon}(h_l)=A_1^{\epsilon}(1+
b_{1,\epsilon}|h_l|^{\frac{8}{15}}+
b_{2,\epsilon}|h_l|^{\frac{16}{15}})~~~.
\en

Differently from the cases studied in~\cite{ch99} keeping only
two terms in the scaling function is enough due to the large errors in the
input data.

The final results are 
\eq
8.090 < |A_1^{\sigma}|< 8.125
\en
\eq
17.0 < |A_1^{\epsilon}|< 18.1
\en
which perfectly agree with the predictions of~\cite{d2001,lm99} reported in 
tab.\ref{tab1}

Then (similarly to what we did in~\cite{ch99}) 
we tried to extract the dominant correction to scaling by assuming
 as fixed inputs
the values  for $A_1^{\Phi}$ reported in~tab.\ref{tab1}
 and performing again the fits.
  The results are:
\eq
0.56 < b_{1,\sigma} < 0.66
\en
\eq
-0.96 < b_{1,\epsilon} < -0.82
\en
As it also happens for the energy itself (see~\cite{ch99}) the 
leading correction to scaling in the energy sector is rather large and negative
in sign.

\end{description}

\begin{table}[h]
\caption{\sl Results of the fits.}
\vskip 0.2cm
\begin{tabular}{|l|l|l|}
\hline
    $h_l$     &   $-A_1^{\sigma}$         & $-A_1^{\epsilon}$    \\
\hline
0.20      &           9.186(1)
&                9.14(1)
\\
0.19       &          9.1185(15) 
 &               9.625(15) 
\\
0.18        &        9.0365(15)  
  &             9.935(15)  
\\
0.17         &       8.9735(15)  
   &            10.24(1) 
\\
0.16          &      8.912(1)  
    &           10.55(1) 
\\
0.15           &     8.843(2)  
     &          10.97(2) 
\\
0.14            &    8.770(4)  
      &         11.44(3) 
\\
0.13             &   8.730(2)  
       &        11.665(10)
\\
0.12              &  8.660(5)  
        &       12.035(10)
\\
0.11    &            8.612(5)  
         &      12.34(4)  
\\
0.10     &           8.562(2)  
          &     12.64(2)  
\\
0.09      &          8.512(2)  
           &    12.95(2)  
\\
0.08       &         8.460(2)  
            &   13.35(5)  
\\
0.075       &        8.435(4)  
             &  13.52(5)  
\\
0.066103019026467 &  8.389(2)  
&   13.83(5) 
\\
0.055085849188723  & 8.337(2)  
 &  14.27(5) 
\\
0.044068679350978  & 8.288(2)  
  & 14.65(5) 
\\
0.033051509513233  & 8.237(4)  
 & 15.15(10) 
\\
0.022034339675489  & 8.206(20) 
  &15.68(15) 
\\
\hline
\end{tabular}
\label{tab3}
\end{table}

\subsection{An attempt to obtain $A_2^{\Phi}$}

It is impossible to extract the $A_2^\Phi$ by directly fitting. Moreover we
cannot improve the situation by using the known continuum limit values of 
$<\Phi>$ and $A_1^\Phi$ due to the large correction to scaling whose uncertainty
is of the same order of magnitude of the terms that we would like to observe.
The only possible way out is to construct a combination of the input data which
exactly eliminates $<\Phi>$ and $A_1^\Phi$. This can be easily done by combining
the values of $\Phi$ measured at three different values of the lattice size
$R_1,R_2,R_3$. The
combination is the following:
\eq
A_2^\Phi=\frac{\pi}{\langle\Phi\rangle_{R=\infty}}
\frac{[\Phi(R_1)-\Phi(R_2)]\Delta_1(1,3) -
[\Phi(R_1)-\Phi(R_3)]\Delta_1(1,2)}
{\Delta_2(1,2)\Delta_l(1,3)-\Delta_2(1,3)\Delta_l(1,2)}
\en
where
\eq
\Delta_l(i,j)\equiv [K_0(m_lR_i-K_0(m_lR_j)]~~~.
\en
The result obtained using as input data those for $h=0.022034...$
for the magnetization are reported in tab.\ref{tab4}. The quality of the result
is rather good and is compatible with the prediction for this constant reported
in tab.\ref{tab1}. Notice however that the quality of the results becomes worse
and worse as $h$ increases (they are already almost completely unstable at
$h=0.055085...$) moreover no result can be obtained for $\epsilon$ even at
$h=0.022034...$. This can be easily understood. It due to the fact that
the absolute value of $<\epsilon>$ is  much smaller  than that of $<\sigma>$ and
as a consequence the relative errors (which are those which generate the
observed instabilities) are larger. We have checked that in our range of values
of $h$ higher contributions in the expansion of eq.(\ref{universal}) 
essentially give no effect.

\begin{table}[h]
\caption{\sl Tentative estimate for $A_2^\sigma$.}
\vskip 0.2cm
\begin{tabular}{|l|l|l|l|}
\hline
    $R_1$ & $R_2$ & $R_3$ &      $-A_2^{\sigma}$    \\
\hline
          15&          16&          17&  -22.3477682808347     \\
          15&          16&          18&  -22.3151718797824     \\
          15&          16&          19&  -22.2853020505359     \\
          15&          16&          20&  -22.2593381627093     \\
          15&          17&          18&  -22.2646875236694     \\
          15&          17&          19&  -22.2261047816289     \\
          15&          17&          20&  -22.1918745162029     \\
          15&          18&          19&  -22.1652771489821     \\
          15&          18&          20&  -22.1211891807013     \\
          15&          19&          20&  -22.0505359976403     \\
          16&          17&          18&  -22.1917546952317     \\
          16&          17&          19&  -22.1416354502675     \\
          16&          17&          20&  -22.0965339005886     \\
          16&          18&          19&  -22.0640813462388     \\
          16&          18&          20&  -22.0064100955253     \\
          16&          19&          20&  -21.9155756522890     \\
          17&          18&          19&  -21.9520757204636     \\
          17&          18&          20&  -21.8778373894290     \\
          17&          19&          20&  -21.7630532199712     \\
          18&          19&          20&  -21.5973234560565     \\
\hline
\end{tabular}
\label{tab4}
\end{table}

\section{Concluding remarks}

Let us briefly summarize our main results.

\begin{itemize}
\item 
Our results nicely agree with the functional form for the finite size behaviour
of the one-point functions 
proposed by Delfino and LeClair-Mussardo. 
We are also able to give
a good estimate of the first correction to scaling terms, both for 
$\langle\sigma\rangle$ and for $\langle\epsilon\rangle$
 and a rough but
reliable estimate of the $A_2^\sigma$  for the
magnetization. 
\item 
 Unfortunately we are not able to discriminate between the two proposals. This
 would require precisions which are definitely outside the range of our
 transfer matrix methods.
\item
The asymptotic values for the energy and the magnetization obtained implementing
eq.(\ref{universal}) are more precise that those that  quoted
in~\cite{ch99} of more than one order of magnitude. 
As the critical point is approached this improvement in precision increases.
\item
The agreement between the present results and those of~\cite{ch99}, besides
being a further strong evidence of the correctness of the proposal
of~\cite{d2001} and \cite{lm99}, also points out the power of
the algorithm proposed
in~\cite{ch99} to deal with the finite size correction. This is an important
result since this algorithm is not constrained to integrable theories

\end{itemize}

It would be interesting to extend this analysis 
to other integrable models which could offer
examples in which it could be simpler to discriminate between the two proposals
of~\cite{d2001} and \cite{lm99} and to compare them with other existing
approaches (say, for instance,~\cite{l01}). 
It would also be important to extend the analysis to
 models which could be more easily
accessible for comparison with experimental results 
(see for instance the results of~\cite{k01,kf01}
on the Haldane-Gapped spin chains).
 At the same time it would be interesting to extend
the present analysis to two-point functions, for which 
very precise numerical results exist for the 2d Ising model in a magnetic
field~\cite{cgm99,cgm2001}. This would be a rather important test since it has
been recently claimed~\cite{s99,cf2002} that the extension to two-point 
functions of the proposals~\cite{d2001} 
and \cite{lm99} could fail in the case of interacting theories.

\subsection*{Acknowledgements}
We thank G.~Delfino and G.~Mussardo for several useful discussions on the
subject.

\end{document}